\documentclass[11pt,twoside]{article}
\usepackage{epsfig}
\usepackage{psfig}
\usepackage{here}
\usepackage{amssymb}
\usepackage{amsmath}
\usepackage[scriptsize]{subfigure}
\usepackage{amsmath}
 
\begin{document}

\begin{center}{\bf\Large Polarized Proton and Deuteron Beams at COSY} 
\end{center}
\vspace{0.5cm}
\begin{center}
   H.~Stockhorst for the COSY spin team\\[0.1cm]
   {\small \em 
         Institut f\"ur Kernphysik, Forschungszentrum J\"ulich, Germany \\          
   }
\end{center}
\vspace{0.5cm}
\begin{center}
 \parbox{0.9\textwidth}{
  \small{
    {\bf Abstract:}\
     Aim of this paper is to present a brief and basic introduction to polarized beams and the behaviour of a particle's spin in an electromagnetic field. It is not intended to give a complete presentation of the spin dynamic theory. Instead the most important features are given without proof and details not absolutely necessary are left out. Spin depolarizing mechanisms are discussed and the cures against depolarizing especially during acceleration as utilized at COSY are outlined.  
A list of references is given at the end to help the reader to find more deep details on spin dynamic theory~\cite{literatura}.
     }
 }
\end{center}

\vspace{0.5cm}
\section{Introduction}
The particle spin is an important feature of nuclei and subnuclear particles. In general, the particle interactions depend to various degrees on their spin states. It is thus not surprising that the experimentalists´ demand on polarized beams at COSY~\cite{cosy} became important rather early. Polarized proton beams at COSY are now routinely delivered to internal and external experiments since 1996. The polarized beams from the ion source are pre-accelerated in the cyclotron JULIC~\cite{beuscher}, injected and accelerated in COSY without significant loss of polarization. Imperfection resonances are increased in strength by means of vertical orbit distortions, leading to a complete polarization reversal (spin flip). Intrinsic resonances are overcome by means of a fast ramping air core quadrupole magnet inducing a rapid change in tune and therefore preservation of the polarization (tune jump).

The polarization during acceleration is observed utilizing parts of the former EDDA~\cite{edda} detector system as polarimeter. Only the fast online polarization measurements during the COSY acceleration cycle made possible by the EDDA setup allows an efficient correction of the up to 16 first order depolarizing resonances encountered.

Polarized deuterons were accelerated for the first time early in 2003 and are also now routinely available for internal and external experiments. Because of the lack of depolarizing resonances in the COSY energy range, preservation of the polarization is less involved for deuterons as compared to protons. However, a polarization measurement technique for vector and tensor polarization has to be developed. Also in this case the EDDA detector turned out to be the most important and efficient tool to measure the polarization even in the accelerating ramp of COSY.

Polarized proton beams with intensities above 10$^{10}$ stored protons with a degree of polarization of up to 0.80 are now routinely delivered to internal experiments~\cite{cosy11-1,cosy11-2}. For external experiments polarized protons can be slowly extracted from COSY via stochastic extraction without loss of polarization by keeping the extraction momentum far enough away from depolarizing resonances and careful adjustment of the betatron tunes during the extraction process.

\section{Polarization}
\subsection{Polarization of spin-1/2 particle beams}

Along a quantization axis $z$ the spin $\vec{S}$ of a proton can only be found in two eigenstates of $\vec{S}_{z}$ with eigenvalues of the spin component $S_{z}=+\hbar/2$ ("up" or "+" state) or $S_{z}=-\hbar/2$ ("down" or "-" state). The degree of polarization is defined as the ratio

\begin{equation}
P~=~\frac{N_{+}-N_{-}}{N_{+}+N_{-}},
\label{eq1}
\end{equation}
where N$_{\pm}$ are the number of particles in the two spin states "up" and "down". For a 100$\%$ polarized beam (P~=~$\pm$1) the spin states of all particles are aligned along a polarization axis. In general three parameters are needed to specify the polarization of spin-1/2 particles. Two for the direction of the quantization axis and one for the ratio $\frac{N_{+}}{N_{-}}$. Thus the polarization of a spin-1/2 particle is a vector characterized by a direction and magnitude. This type of polarization is therefore called $vector$ $polarization$. The polarization of a bunch of particles is found by an ensemble average.

\subsection{Polarization of spin-1 particle beams}
\begin{figure}[H]
\begin{center}
\parbox{0.4\textwidth}
   {\epsfig{file=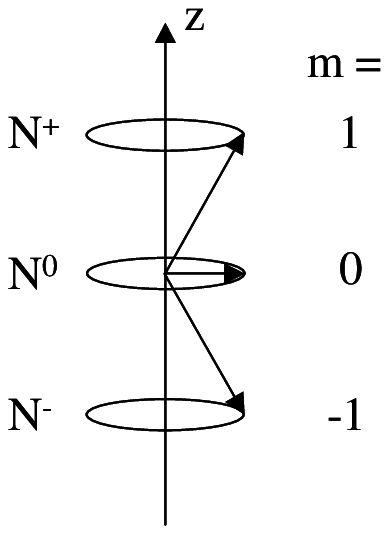,scale=0.7}}
\hfill
\parbox{0.5\textwidth}
  {\caption{\label{fig1} \small Spin states for a spin-1 particle. 
   }}
\end{center}
\end{figure}

For a spin-1 system, e.g. deuterons, there are three spin states along a quantization axis as depicted in figure~\ref{fig1}.

Particles having their spin in a $m=0$ state have zero vector polarization $but$ they are still aligned, i.e. are polarized. Thus the polarization of spin-1 particles is characterized by vector polarization $and$ an alignment. This is called $tensor$ $polarization$. The vector polarization is given by the ratio
 
\begin{equation} 
P_{z}=\frac{N_{+}-N_{-}}{N_{+}+N_{0}+N_{-}},
\label{eq2}
\end{equation}
where $N_{+}$, $N_{0}$, $N_{-}$  particles are aligned along the axis with $m=1, 0, -1$ respectively. The tensor polarization is determined from 
\begin{equation}
P_{zz}=\frac{N_{+}+N_{-}-2N_{0}}{N_{+}+N_{0}+N_{-}},
\label{eq2}
\end{equation}              
with $-2\leq P_{zz}\leq 1$.

There are several possibilities for vector and tensor polarization. If all states were equally populated the polarization would be zero. Some special polarization states are: 

\begin{itemize}
\item{maximum vector polarization $P_{z}=\pm1$ with tensor polarization $P_{zz}~=~1$}
\item{pure vector polarization ($P_{zz}=0$) with $P_{z}=\pm 2/3$}
\item{pure tensor polarization with $P_{zz}=-2$ or $1$ and $P_z=0$}
\item{$P_{z}=\pm 1/3$ with $P_{zz}=-1$}
\end{itemize} 

The polarized source of COSY is capable to deliver these states as required from the experiment. 
 
The polarization of a particle $\vec{P}$ is the quantum average of the spin operator $\vec{S}$, 
$\vec{P}=\frac{2}{\hbar}<\psi|\vec{S}|\psi>$. According to the Ehrenfest theorem the evolution of the polarization vector (which is called $\vec{S}$ in the following) is governed by a classical equation of motion.

\section{Spin Precession in an Electromagnetic Field}
\subsection{Non-relativistic particles}
\begin{figure}[H]
\parbox{0.4\textwidth}
   {\epsfig{file=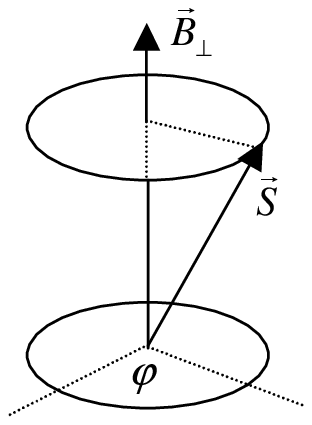,width=0.4\textwidth}}
\hfill
\parbox{0.5\textwidth}
  {\caption{\label{fig2} \small Precession of the spin vector around the magnetic field. 
   }}
\end{figure}

For non-relativistic non-radiating particles in a magnetic field $\vec{B}$ 
the spin motion in the particles' rest frame is governed by the equation
\begin{equation}
\frac{d\vec{S}}{dt}=\vec{\mu}\times \vec{B}
\label{eq3}
\end{equation}
that gives the time change of the spin as the torque on the magnetic moment $\vec{\mu}$
exerted by the magnetic field. The magnetic moment $\vec{\mu}$ is associated to the spin by
\begin{equation}
\vec{\mu} = g\frac{e}{2m_0}\vec{S}
\label{eq4}
\end{equation}
with the rest mass $m_0$ and the elementary charge $e$. The gyromagnetic ratio $g$
is 2 for a point-like spin-1/2 particle. For real particles its deviation from $2$ 
is expressed by the gyromagnetic anomaly $G=(g-2)/2$. For protons this 
value is $G=1.79$ and for deuterons $G=-0.143$.
If we insert the magnetic moment into eq.~\ref{eq3} the equation of motion of the spin vector is found
\begin{equation}
\frac{d\vec{S}}{dt}=\vec{S}\times \vec{\Omega}_{L}
\label{eq5}
\end{equation}
around the direction of the rotation vector (angular velocity)
\begin{equation}
\vec{\Omega}_{L}=\frac{ge}{2m_0}\vec{B}
\label{eq6}
\end{equation}
which is collinear to the magnetic field, see figure 2.  
The magnitude of the rotation vector gives the time rate of change $d\phi /dt$
of the angle $\phi$ covered by the spin projetion onto the plane perpendicular to the magnetic field,
\begin{equation}
\frac{d\phi}{dt}=|\vec{\Omega}_{L}|,
\label{eq7}
\end{equation} 
as shown in figure~\ref{fig2}.

\subsection{Relativistic particles}
\begin{figure}[H]
\parbox{0.4\textwidth}
   {\epsfig{file=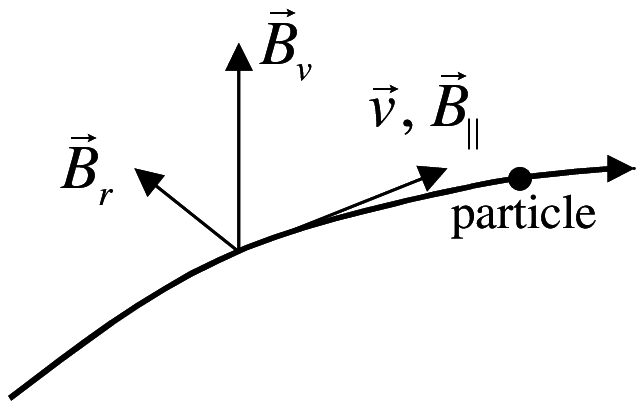,width=0.4\textwidth}}
\hfill
\parbox{0.5\textwidth}
  {\caption{\label{fig3} \small Decomposition of the magnetic field into a component parallel $\vec{B}_{\parallel}$ to the beam's velocity and orthogonal to it $\vec{B}_\perp$, in this case a radial and vertical component. 
   }}
\end{figure}

So far we considered a particle at rest. However in COSY particles are moving with 
a speed close to the velocity of light and also experience an acceleration when they are bent by the
dipole magnets. It turns out that eq.~\ref{eq5} is then still valid albeit with a different rotation vector
if the electromagnetic fields in the laboratory frame are properly Lorentz transformed into a co-ordinate 
system that moves with the particle around the ring (moving frame). If for simplicity the particles 
move in the horizontal plane they experience an acceleration that is perpendicular to the velocity 
and the vertical magnetic field (dipole fields of COSY). This is the reason of the Thomas precession 
which contributes an angular velocity in the moving frame. Thus the angular velocity $\vec{\Omega}$ in eq.~\ref{eq5} will 
contain two parts, one component results from the Lorentz transformation and the other one from the 
Thomas precession term. The correct form of eq.~\ref{eq5} for moving particle bears the name from $Thomas$, 
$Bargman$, $Michel$ and $Telegdi$ $(Thomas-BMT$ $equation)$ and is given by
\begin{equation}
\frac{d\vec{S}}{dt}=\vec{S}\times \vec{\Omega}_{BMT}
\label{eq8}
\end{equation}
with the angular velocity vector
\begin{equation}
\vec{\Omega}_{BMT}=\frac{e}{m_{0}\gamma}[(1+\gamma G)\vec{B}_{\perp}+(1+G)\vec{B}_{\parallel} 
-(\gamma G+\frac{\gamma}{1+\gamma})\vec{\beta}\times\frac{\vec{E}}{c}]
\label{eq9}
\end{equation}
in which the magnetic field $\vec{B}$ (see figure~\ref{fig3}) in the laboratory frame has been decomposed into a componet
$\vec{B}_{\parallel}$ parallel to the particle velocity $\vec{\beta}=\vec{v}/c$
and perpendicular to it, $\vec{B}_{\perp}$. The electric field is $\vec{E}$
and the kinematic factor is given by $\gamma=\frac{1}{\sqrt{1-\beta^{2}}}$.

The spin precession has the following properties:
\begin{itemize}
\item{An electric field ${\bf E}$ is equivalent to a magnetic field ${\bf B}$=${\bf E}/c$. An electric field of $3\cdot 10^{8}$V/m has the same strength as a magnetic field of 1~T. In general no such electric field is present. 
{\it In the following electric fields are therefore neglected}.}
\item{  Comparison of longitudinal ($\parallel$) and transverse ($\perp$) magnetic fields:
According to eq.~\ref{eq7} and~\ref{eq9} the spin vector will rotate in longitudinal fields by
$d\phi=\Omega_{BMT}dt$ in the time interval $dt$. Using $ds/dt=\beta c$ we see that the spin rotates by
$d\phi=\Omega_{BMT}ds/(\beta c)$ while the particle moves the distance $ds$. Thus the total spin rotation in a longitudinal field is
\begin{equation}
\Delta\phi=\frac{e}{p}(1+G)\int{B_\parallel ds},
\label{eq10}
\end{equation}
where we have used that the momentum of the particle is given by $p~=~m_0\beta\gamma c$.
Similarly, for transverse fields the spin rotation is determined from
\begin{equation}
\Delta\phi=\frac{e}{p}(1+\gamma G)\int{B_\perp ds}=\frac{e}{m_0\beta\gamma c}(1+\gamma G)\int{B_\perp ds}.
\label{eq11}
\end{equation}
The spin rotation by longitudinal fields becomes less effective at large momenta and the spin rotation by transverse fields becomes nearly independent of momentum ($\gamma G \ll 1$).}
\end{itemize}
From eq.~\ref{eq10} and~\ref{eq11} we can directly calculate that 
\begin{itemize}
\item{to rotate the spin of 3.3~GeV/c protons around the $longitudinal$ axis by an angle $\pi$ 
a $longitudinal$ field with 12.4~Tm is needed at COSY and for deuterons: 40.3~Tm (!!)} 
\item{to rotate the proton spin at 3.3~GeV/c in the $moving$ $frame$ by an angle $\pi$ around 
the $vertical$ axis one needs a vertical magnetic field with
$\int{B_\perp ds}=\pi\frac{m_0\beta c}{eG}=5.3~Tm$, and for deuterons~$\approx~118~Tm$.}
\end{itemize}
Thus the deuteron´s spin is about 20 times harder to manipulate with magnetic fields. On the other hand their spin direction is less sensitive to field errors.

\section{Spin Precession and Resonances in a Circular Particle Accelerator}
We first consider an ideal accelerator or storage ring in which all components (dipoles quadrupoles etc.) lie in the horizontal plane. The particles will then also move in the horizontal plane on the ideal (circular) path guided by the vertical dipole fields. The particle frame rotates with angular velocity (cyclotron frequency)
\begin{equation}
\vec{\Omega}_C=\frac{e}{m_0\gamma}\vec{B}_\perp.
\label{eq12}
\end{equation}
The spin will precess around the vertical axis and remains constant in the straight sections as long as the particle passes the quadrupoles on axis. The difference in frequency of spin and velocity precession, eq.~\ref{eq9} and eq.~\ref{eq12}, yields
\begin{equation}
\vec{\Omega}=\vec{\Omega}_{BMT}-\vec{\Omega}_C=\gamma G\vec{\Omega}_C.
\label{eq13}
\end{equation}
The {\it number of spin precession during one turn} in the ring follows from eq.~\ref{eq13}
and is called {\it spin tune}
\begin{equation}
\nu=\gamma G.
\label{eq14}
\end{equation}
From eq.~\ref{eq9} it is clear that the vertical spin component is preserved while the two other components rotate in the moving frame with the angular frequency given by eq.~\ref{eq13} around the vertical axis. As the particle completes one revolution, the coordinate system rotates by $2\pi$ and the spin has precessed around the vertical axis by $2\pi \gamma G$. As a result, in a ring where particles experience only vertical magnetic fields the only stable (or invariant) spin direction is along the vertical axis. The spin rotation around the vertical axis after one turn in the ring can be expressed by the matrix

\begin{equation}
\left(
\begin{array}{ccc}
      \cos(2\pi\gamma G) & 0 & \sin(2\pi\gamma G) \\
        0 & 1 & 0 \\
     -\sin(2\pi\gamma G) & 0 & \cos(2\pi\gamma G) 
\end{array}
\right)
\label{eq15}
\end{equation}

Now consider that at one location in the ring there is a static $radial$ magnetic field component (along the $x$-axis) 
as shown in figure~\ref{fig3}. When the particle passes this disturbing magnetic field the spin is rotated around the $x$-axis
by an angle ( given by eq.~\ref{eq11}). The spin rotation around the $x$-axis due to the disturbing field is then given by the matrix

\begin{equation}
\left(
\begin{array}{ccc}
      1 & 0 & 0 \\
      0 & \cos\theta & \sin\theta \\
      0 & -\sin\theta & \cos\theta 
     
\end{array}
\right)
\label{eq16}
\end{equation}

The total spin rotation matrix for one turn is then given by the product $M$ of eq.~\ref{eq15}
and eq.~\ref{eq16}. The stable spin direction outside the disturbing field region can be found by determining the eigenvectors of 
$M$ with eigenvalue 1, i.e. solving the equation $M\vec{u}=\vec{u}$. One finds that the vertical spin component is:
\begin{equation}
y=\frac{S_y}{\sqrt{1+\tan^{2}(\theta/2)/\sin^{2}(\pi\gamma G)}}.
\label{eq17}
\end{equation}
Clearly, $y=S_y$  if there is no disturbing field, i.e. $\theta=0$. However polarization along the vertical axis is completely lost if the spin tune equals an integer, $\nu=\gamma G=k\in \mathbb{N} $. One can say that the spin motion is in $resonance$
with the radial magnetic field: every time the particle passes the field the spin vector is bent more and more away from the vertical axis until polarization is lost. Resonances of this type are called $imperfection$ $resonances$ or $integer$ $resonances$.
They arise e.g. if magnets are slightly misaligned or if there are vertical orbit distortions. The particles then experience radial magnetic fields (e.g. from vertical focussing quadrupoles).
Thus it is impossible to operate the accelerator with a polarized beam at energies where imperfection resonances occur. However they can be crossed with (almost) no polarization loss during acceleration as is discussed below.

\begin{figure}[H]
\parbox{0.4\textwidth}
   {\epsfig{file=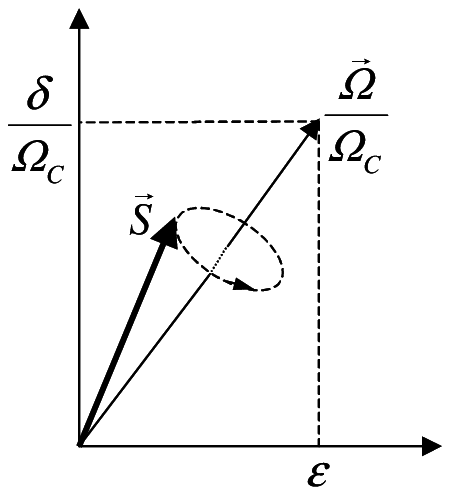,width=0.4\textwidth}}
\hfill
\parbox{0.5\textwidth}
  {\caption{\label{fig4} \small Explanation see text. 
   }}
\end{figure}

The term "resonance" might become more clear in another view: Each time the particle passes the disturbing radial field the spin may point in a different direction with respect to the magnetic field if the spin tune is not an integer. One can think of a rotation of the radial magnetic field around the vertical axis with resonance frequency $\nu_R\Omega_C$. 
While the spin is rotating around the vertical axis with spin tune $\nu$ (not an integer) it rotates with frequency
$\delta=(\nu-\nu_R)\Omega_C$ in the frame of the disturbing field as shown in figure~\ref{fig4}.
If the spin tune is not an integer these rotations are out of phase and the disturbing effect of the field averages out. No polarization loss occurs.

\begin{figure}[H]
\parbox{0.4\textwidth}
   {\epsfig{file=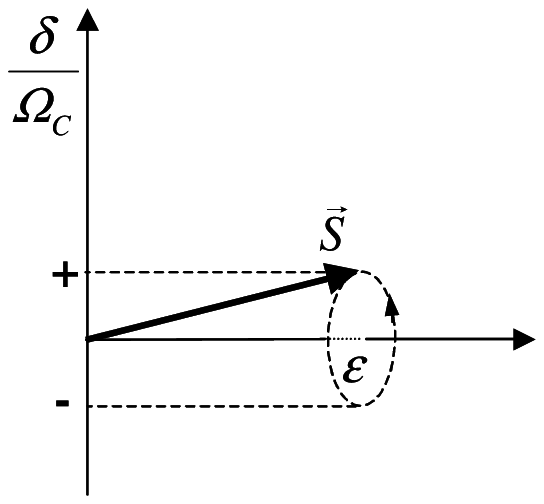,width=0.4\textwidth}}
\hfill
\parbox{0.5\textwidth}
  {\caption{\label{fig5} \small On resonance the vertical polarization is lost. 
   }}
\end{figure}

The radial component is according to eq.~\ref{eq9} given by
\begin{equation} 
\vec{\Omega}_\perp=\frac{e}{m_0\gamma}(1+\gamma G)\vec{B}_\perp=\epsilon\vec{\Omega}_C, 
\end{equation} 
where $\epsilon$ is proportional to the $resonance$ $strength$. 
It is clear that the larger the field is the larger the resonance strength will be. Large vertical closed orbit errors enhances the imperfection resonance strength. The resonance strength can also be interpreted as the $width$ $of$ $the$ $resonance$ 
since in the tune interval $(\nu_R-\epsilon,\nu_R+\epsilon)$ the angle of $\vec{\Omega}$
with vertical axis is larger then 45 degrees.
On top of the resonance, i.e. $\delta=0$,
as shown in figure~\ref{fig5}, the spin rotates around the radial direction and on average the vertical component is zero and the vertical polarization is lost.

So far we assumed that the particles move, apart from the straight sections, on a circle in the horizontal plane. If the particles execute $horizontal$
betatron oscillations the spin rotation is no longer given by the rotation matrix eq.~\ref{eq15}
since the path length during one turn of a particle now depends on the betatron amplitude and betatron tune $Q_x$.
The spin precession is $modulated$
by the betatron motion and the spin tune has two additional values which are in the case of COSY $\gamma G\pm(Q_x-2)$.
A $horizontal$ $spin$ $resonance$ can occur if $\gamma G\pm(Q_x-2)$ equals an integer.
These resonance are not important (except when there is coupling between the horizontal and vertical betatron motion) at COSY with its 
\underline{vertically} polarized beams since the horizontal betatron motion is driven by vertical magnetic fields of the quadrupoles (see eqs.~\ref{eq8} and~\ref{eq9}).
In addition to the horizontal betatron motion the particles also execute a vertical betatron motion with betatron tune $Q_y$
induced by the \underline{radial magnetic fields} 
of the vertical focussing quadrupoles. In this case eq.~\ref{eq16} is no longer valid since, besides a possible static radial magnetic field, the spin precession experiences, or is driven by a force that is evoked by the vertical betatron motion.
Similarly, a resonance occurs if $\gamma G\pm(Q_y-2)$ equals an integer.
These resonances are named $vertical$ $spin$ resonances or more frequently, $intrinsic$ $resonances$.
Their strength is proportional to the square root of the vertical beam emittance (proportional to beam size). Resonances of this type are dangerous in COSY because they produce a severe polarization loss if they are not avoided or compensated. As in the case of imperfection resonances the spin resonances driven by the betatron motion are energy dependent but in addition depend also on the vertical tune of the machine.
Among these resonances there are higher order resonances as well those that can be excited by the synchrotron motion of the particles as induced by an rf-cavity (bunched beam). These cases play however a minor role at COSY and will therefore not be discussed here. There are also resonances that can be excited by longitudinal magnetic fields (solenoids). For this type of resonance the reader is referred to the literature. 

\section{Resonance Crossing}
Suppose we are away from a single resonance. Then $\delta=(\nu-\nu_R)\Omega_C\neq 0$.
If the beam is accelerated linearly through the resonance one can write $\nu=\nu_R+\alpha\Theta$
for the linear variation of the tune with azimuth $\Theta$. For the spin rotation frequency on resonance we have
$\Omega_\perp=\epsilon\Omega_C$ which shows that the change in the spin rotation angle within the "time"
$\Delta\Theta$ is $\Delta\phi=\epsilon\Delta\Theta$. Since the "time" needed to cross the resonance is
$\Delta\Theta=2\epsilon/\alpha$
the spin rotation angle changes by $\Delta\phi=2\epsilon^{2}/\alpha$ during resonance crossing. Thus we can distinguish two regimes:
\begin{itemize}
\item{{\bf Fast crossing} ($\Delta\phi\ll 1$) means either a small resonance strength or a large crossing speed.
There is no time to bent away the spin $\vec{S}$ from the vertical and polarization is conserved.}
\item{{\bf Slow crossing} ($\Delta\phi\gg 1$) means either a large resonance strength or a slow crossing speed.
The spin vector rotates much faster around the rotation vector $\vec{\Omega}$ than the latter moves.
The spin vector $\vec{S}$  follows the motion of $\vec{\Omega}$  and is completely reversed. 
The polarization has changed its sign but there is no depolarization ($adiabatic$ $spin$ $flip$).}
\end{itemize}

Between these two cases polarization may be completely lost. A quantitative estimate of the final vertical spin component 
$S_y$ compared to the initial value before crossing the resonance is given by the Froissart-Stora formula
\begin{equation}
\frac{S_{y,initial}}{S_{y,final}}=2e^{\frac{\pi\epsilon^{2}}{2\alpha}}-1,
\label{eq18}
\end{equation}
which includes the two cases of fast crossing, $S_{y,final}=+S_{y,initial}$, and adiabatic spin flip,
$S_{y,final}=-S_{y,initial}$.

\section{Polarized Proton and Deuteron Beams at COSY}
\subsection{Proton beams}

At COSY vertically polarized proton beams are available in the momentum range up to 3650~MeV/c for $internal$ and $external$ experiments.
At flat top about $1.0\cdot 10^{10}$ protons with a polarization of more than 80$\%$
can be achieved. During acceleration to flat top energy up to sixteen depolarizing resonances have to be cured for polarized protons.
The following graphics shows the imperfection and intrinsic resonances in the momentum range of COSY. In this figure the imperfection resonances appear as horizontal lines since they are independent of tune.
The horizontal axis is the fractional vertical tune $q_y$ ($Q_y$=$integer$ $part$ + $q_y$,  for COSY the integer part always equals 3).

\begin{figure}[H]
\vspace{-1.8cm}
\parbox{1.0\textwidth}
   {\epsfig{file=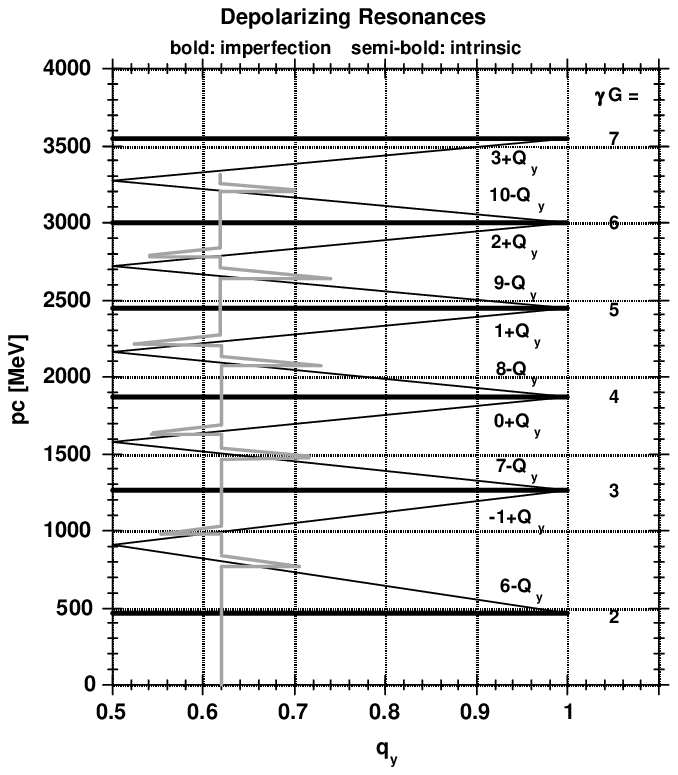,width=1.0\textwidth}}
\hfill
\vspace{-1.8cm}
  \caption{\label{fig6} \small Depolarizing resonances during acceleration 
   of protons in COSY as a function of the vertical fractional tune. Bold: imperfection resonances, semi-bold: intrinsic resonances.
   The shaded zigzag line at $q_y=0.62$ symbolizes the change in working point (not in scale) during acceleration of the beam (tune jump).  
   }
\end{figure}

\begin{figure}[H]
\vspace{-1.0cm}
\hspace{1.3cm}
\parbox{1.0\textwidth}
   {\epsfig{file=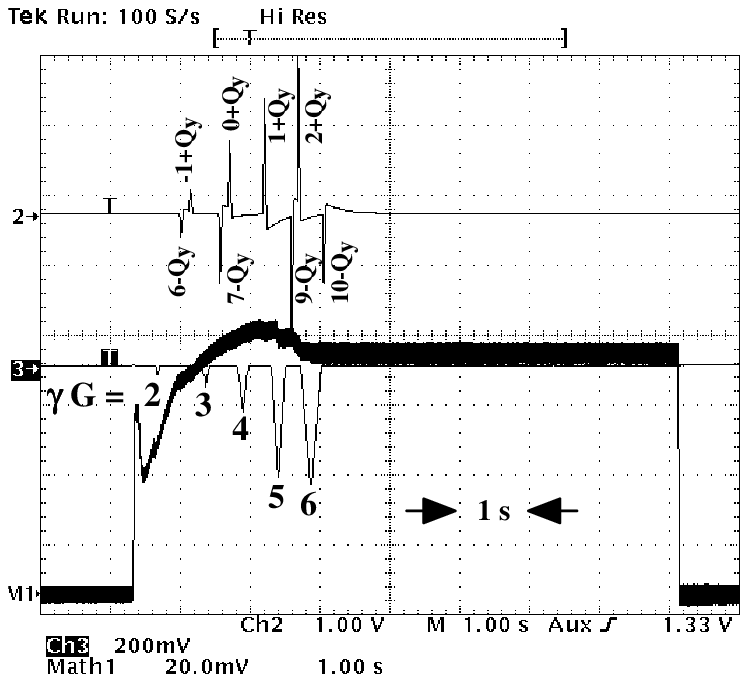,width=0.7\textwidth}}
\hfill
  {\caption{\label{fig7} \small Trace M1 shows the beam current versus time measured with a beam current transformer (BCT). Trace 3 represents the current of the vertical steerer magnet that excites a total spin flip at the imperfection resonances. The current of the fast quadrupole is given by trace 2. 
   }}
\end{figure}

There are six imperfection resonances. A loss of polarization at these resonances is avoided by artificially increasing the resonance strength, such that the polarization direction is completely reversed (adiabatic spin flip, slow crossing) when protons are accelerated across the resonance. A vertical correction dipole located at a position with a large vertical beta-function is used to increase the orbit distortion that enhances the resonance strength. One method to conserve polarization at up to ten intrinsic resonances is a rapid vertical tune change (fast crossing). For this a fast tune jump system consisting of one air coil quadrupole with a length of 0.6 m and a gradient of 0.43 T/m is installed in one arc of COSY.
It allows a rapid tune change of about 0.06 within 10~$\mu s$. Double crossing of resonances is avoided by a slow fall time of 40~ms.
Polarization and particle losses due to an emittance increase can be kept low during acceleration if the beam position is carefully aligned in the acceleration ramp.
In addition electron cooling at injection energy of COSY has been successfully applied to reduce the beam emittances in the transverse as well as in the longitudinal phase space. This increases not only the fast crossing (chapter 5) by reducing the resonance strength but also lowers beam losses visibile in figure 5 caused by the fast tune jumps. At injection the vertical tune is below 3.66. Therefore it is fixed close to 3.62 during acceleration in order to have enough time for consecutive tune jumps (figure~\ref{fig6}).
The dynamic tune measurement allows adjusting the tune during acceleration as well as the time and amplitude of the tune jumps. 
Figure~\ref{fig7} shows online measurements of the vertical steerer and fast quadrupole currents in comparison with the beam current (BCT). The beam is accelerated in 2.6 s from injection, 0.294 GeV/c, up to 3.1 GeV/c. Particle losses due to the steerer and fast quadrupole actions are less than 10$\%$. Electron cooling was not applied in this case. In this example $4\cdot 10^9$ polarized protons are accelerated to flat top.

\begin{figure}[H]
\vspace{-1.0cm}
\parbox{1.0\textwidth}
   {\epsfig{file=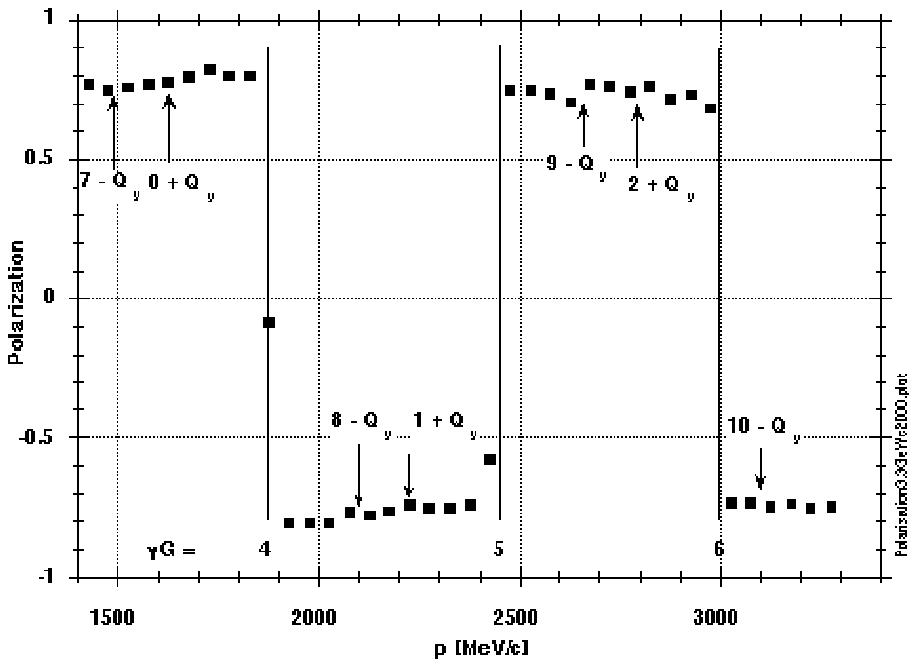,width=0.9\textwidth}}
\hfill
\vspace{-0.4cm}
  {\caption{\label{fig8} \small Online polarization measurement starting about 1.3 GeV/c. The arrows mark the intrinsic resonances which are compensated with almost no polarization loss by fast tune jumps.  A total spin flip is created with a vertical steerer for the imperfection resonances. 
   }}
\end{figure}

Figure~\ref{fig8} shows a polarization measurement versus momentum during acceleration to 3.3 GeV/c. The initial polarization is 80$\%$
and drops to 75$\%$ at flat top momentum. The loss of about 6 
resonance.

On-line polarization measurements are carried out during acceleration with the high precision detector EDDA, designed to measure pp-scattering excitation functions during the acceleration of the COSY beam. This detector is a double layered scintillation hodoscope and in conjunction with a CH$_{2}$ or C fiber target well suited to function as a polarimeter. The great advantage of this online measurement is that the behavior of the polarization at every resonance is clearly visible in only a few ten minutes, depending on the desired measurement statistics. 

\subsection{Deuteron beams}

As was pointed out in chapter 3.2 the deuteron´s spin is about 20 times harder to manipulate with magnetic fields. Due to the deuteron´s
$G$ factor the lowest imperfection resonance that can occur is $\gamma G=7$ corresponding to a momentum
$p=13$~GeV/c which is well above COSY´s working range. Furthermore there are no intrinsic resonances at COSY in the normal tune range
$3.5<Q_y<3.65$.
Stochastic extraction to the experiment GEM~\cite{gem} of a tensor and vector polarized deuteron beam at 2.4~GeV/c flat top momentum has been carried out successfully. 

\vspace{0.8cm}
\hspace{-0.6cm}{\Large\bf Acknowledgment}
\vspace{0.5cm}

I would like to thank the organizers around P. Moskal to invite me for this talk. It was not only a great pleasure to participate in this well prepared and stimulating workshop but also I enjoyed the kind hospitality very much, at the workshop in the impressive premises of the Collegium Maius building and in the splendid neighborhood of Cracow.

\end{document}